\documentclass{PoS}
\usepackage{amsmath}
\usepackage[normalem]{ulem}
\usepackage{tikz}
\usetikzlibrary{arrows}
\tikzstyle{n-style}=[draw, circle, black, fill=gray!20]

\title{Finite size effects on cumulants of the critical mode}

\ShortTitle{Finite size effects on cumulants of the critical mode}

\author{M.~Agah~Nouhou\\
        SUBATECH UMR 6457 (IMT Atlantique, Universit\'e de Nantes, IN2P3/CNRS), 4 rue Alfred Kastler, 44307 Nantes, France\\
        E-mail: \email{mouss1986jan@gmail.com}}
\author{\speaker{M.~Bluhm}\\
        Institute of Theoretical Physics, University of Wroclaw, PL-50-204 Wroclaw, Poland\\
        SUBATECH UMR 6457 (IMT Atlantique, Universit\'e de Nantes, IN2P3/CNRS), 4 rue Alfred Kastler, 44307 Nantes, France\\
        E-mail: \email{marcus.bluhm@subatech.in2p3.fr}}
\author{A.~Borer\\
        SUBATECH UMR 6457 (IMT Atlantique, Universit\'e de Nantes, IN2P3/CNRS), 4 rue Alfred Kastler, 44307 Nantes, France\\
        E-mail: \email{anna.borer@etu.univ-nantes.fr}}
\author{M.~Nahrgang\\
        SUBATECH UMR 6457 (IMT Atlantique, Universit\'e de Nantes, IN2P3/CNRS), 4 rue Alfred Kastler, 44307 Nantes, France\\
        E-mail: \email{marlene.nahrgang@subatech.in2p3.fr}}
\author{T.~Sami\\
        SUBATECH UMR 6457 (IMT Atlantique, Universit\'e de Nantes, IN2P3/CNRS), 4 rue Alfred Kastler, 44307 Nantes, France\\
        E-mail: \email{taklit.sami@subatech.in2p3.fr}}
\author{N.~Touroux\\
        SUBATECH UMR 6457 (IMT Atlantique, Universit\'e de Nantes, IN2P3/CNRS), 4 rue Alfred Kastler, 44307 Nantes, France\\
        E-mail: \email{nathan.touroux@etu.univ-nantes.fr}}

\abstract{In this work we study the temperature dependence of the equilibrium variance of critical fluctuations near the QCD critical point. In particular, we take the finite size of the fireball created in heavy-ion collisions into account and systematically obtain corrections to the leading-order result. We find that not only is the variance globally reduced in a finite size system, but for certain combinations of parameters a two-peak structure can develop for temperatures near the critical point.}

\FullConference{Corfu Summer Institute 2018 "School and Workshops on Elementary Particle Physics and Gravity"\\
		(CORFU2018)\\
		31 August - 28 September, 2018\\
		Corfu, Greece}

\begin{document}

%%%%%%%%%%%%%%%%%%%%%%%%%%%%%%%%%%%%%%%%%%%%%%%%%%%%%%%%%%%%%%%%%%%%%%%%%%%
\section{Introduction}
%%%%%%%%%%%%%%%%%%%%%%%%%%%%%%%%%%%%%%%%%%%%%%%%%%%%%%%%%%%%%%%%%%%%%%%%%%%
The search for the conjectured chiral critical point in the QCD phase diagram is at the heart of a number of ongoing experimental programs: HADES at the GSI Helmholtz Center, NA61/SHINE at the CERN-SPS and the RHIC Beam Energy Scan (BES). By means of relativistic heavy-ion collisions at different geometries, system sizes and beam energies, $\sqrt{s}$, signals for this intriguing landmark are hoped to be found. A prominent signature for its existence is the expected non-monotonic $\sqrt{s}$-dependence of event-by-event fluctuations of measured particle multiplicities related to the conserved charges of QCD~\cite{Stephanov:1998dy,Stephanov:1999zu}. Published data on net-proton, net-electric charge and net-kaon fluctuations~\cite{Adamczyk:2013dal,Adamczyk:2014fia,Adamczyk:2017wsl} are, however, so far inconclusive. Near the critical point, fluctuations of the critical mode are expected to become large as a consequence of the associated increase of the correlation length $\xi$. For a static, equilibrated medium of infinite size higher-order cumulants of these fluctuations were found to diverge at the critical point as $\langle\sigma^n\rangle_c\sim\xi^{5n/2-3}$~\cite{Stephanov:2008qz}. Moreover, a pronounced behavior with $\sqrt{s}$ including wiggles and sign changes might even be expected in certain cumulant ratios~\cite{Asakawa:2009aj,Stephanov:2011pb,Bluhm:2016trm}. 

In reality, the explosively expanding matter created in relativistic heavy-ion collisions is neither infinite nor long-lived. Non-equilibrium, dynamical effects can therefore play a significant role~\cite{Nahrgang:2016ayr} and exact charge conservation in the finite size system might substantially affect our infinite volume expectations. The dynamical effects such as critical slowing down can limit the actual growth of $\xi$~\cite{Berdnikov:1999ph} and lead to the retardation of expected signals~\cite{Herold:2016uvv, Mukherjee:2015swa}. Other late stage, hadronic phase processes such as resonance decays~\cite{Bluhm:2016byc} or isospin randomization~\cite{Kitazawa:2012at} may also reduce the impact of the critical point on observables quantitatively. In comparison to the experiments, thus, realistic simulations embedding the fully coupled critical dynamics are necessary. Major steps in this direction were reported recently in~\cite{Nahrgang:2018afz,Bluhm:2018qkf}. In~\cite{Nahrgang:2018afz}, the diffusive dynamics of net-baryon density fluctuations was studied. In this work, the growth of the correlation length and the scaling behavior of the cumulants with $\xi$ was found to be affected by the finiteness of the system and exact charge conservation during its dynamical evolution.

In this proceeding, we want to discuss the influence of the finite size of the system on the variance of the critical mode fluctuations and its scaling behavior with $\xi$. The infinite volume $V$ expectations may be seen as leading-order results in an $\xi^3/V\ll 1$ expansion. In a realistic setting, however, for a heavy-ion collision probing the region near the QCD critical point higher-order corrections might become non-negligible. This is in particular true as experimentally fluctuations are measured only in a limited rapidity window of the entire fireball.

\section{Finite size effects on the variance}
%%%%%%%%%%%%%%%%%%%%%%%%%%%%%%%%%%%%%%%%%%%%%%%%%%%%%%%%%%%%%%%%%%%%%%%%%%%
In this work we discuss a three-dimensional system of finite size $V=LA$, where we decouple the longitudinal fluctuations from the transverse extension $A$ of the system. This approximation is motivated by the dynamics of a heavy-ion collision which is highly anisotropic. 
 
 %Consider simplification: quasi-1D system, this is experimentally motivated by highly anisotropic situation in a heavy-ion collision and the fact that fluctuations are measured in a given rapidity window

% ... also to motivate the decoupling with transverse area $A$ (which I might set small or not) have to say that we study the simplified case where no correlations develop in the transverse direction or which we ignore for the moment ...
 
% show calculation for second order cumulant, introduce the propagator, in quasi-1D the analog to $\epsilon$ is ... 
 
 %\sout{start with the zero-mode calculation a la [Stephanov]}
 
%% \sout{show relevant diagrams (maybe use texmex (?) as Anna did?)}
 
% correct this: $\sigma$ versus $\Delta\sigma$! Since $\sigma$ is the field, $(\Delta\sigma)^2$ should be the variance, no?! Well $\sigma$ could also be the difference already!
 
% result for $\langle\delta\sigma^2\rangle$ leading order and next to leading order 
 
% \sout{Do I need to specify how a cumulant is defined?} 
 
% Note: effectively this looks like an expansion with increasing powers of $\xi^2/A$! That the third-order cumulant (not discussed here!) starts in LO with a strange power is also observed in\cite{Mukherjee:2015swa}, this seems also to be true for the kurtosis...Check!!!
 
We start our study of finite size effects on the variance of the order parameter $\sigma$ by defining the thermodynamic potential as 
\begin{equation}
 \label{equ:Omega}
  \Omega[\sigma]=\int_V {\rm d}^3x \left(\frac{1}{2}(\vec{\nabla}\sigma)^2+\frac{m^2}{2}\sigma^2+\frac{\lambda_3}{3}\sigma^3+\frac{\lambda_4}{4}\sigma^4+\frac{\lambda_6}{6}\sigma^6\right)
\end{equation}
for a given $V$. Such a form of the potential including a term of order $\sigma^6$ was employed in the recent study~\cite{Nahrgang:2018afz}. The probability distribution in the absence of interactions is given by ${\cal P}_0[\sigma]\sim\exp(-\Omega_0[\sigma]/T)$, where $\Omega_0$ follows from Eq.~(\ref{equ:Omega}) for all $\lambda_i=0$. The free propagator reads 
\begin{equation}
 \label{equ:propagator}
  \langle\sigma_{{k}}\sigma_{-{k}}\rangle=\frac{T}{V}\left({k}^2+m^2\right)^{-1}
\end{equation}
and the corresponding 2-point correlation function connected by Fourier transform is 
\begin{equation}
 \label{equ:CorrFunc1D}
  C_0(x_1,x_2) = \frac{T\xi}{2A} \exp(-|x_1-x_2|/\xi) 
\end{equation}
% \label{equ:CorrFunc3D}
%  C_0({x_1},{x_2})=\frac{T}{4\pi |\vec{x_1}-\vec{x_2}|}\exp(-|\vec{x_1}-\vec{x_2}|/\xi)
% \end{equation}
with correlation length $\xi=1/m$. The coupling coefficients $\lambda_i$ may themselves be functions of $\xi$ and can be obtained from the corresponding universality class. QCD at finite net-baryon density falls into the universality class of the 3d-Ising model. Thus they may be defined as $\lambda_3=\tilde\lambda_3 T(T\xi)^{-3/2}$, $\lambda_4=\tilde\lambda_4 (T\xi)^{-1}$ and $\lambda_6=\tilde\lambda_6/T^2$ with dimensionless coupling coefficients $\tilde\lambda_i$. These and the scaling of the parameters with $\xi$ are universal for the 3d-Ising model, where small corrections from the anomalous dimension $\eta\ll 1$ are neglected. To arrive at Eq.~\eqref{equ:CorrFunc1D} we have scaled out the transverse area $A=V/L$ in the thermodynamic potential Eq.~(\ref{equ:Omega}) 
\begin{equation}
 \label{equ:Omega1D}
  \Omega[\sigma] = A\int_L {\rm d}x \left(\Omega_0[\sigma]+\Omega_{\rm int}[\sigma]\right) \,,
 % \omega_0[\sigma] & = \frac{1}{2}(\vec{\nabla}\sigma)^2+\frac{m^2}{2}\sigma^2\,,\\
 % \omega_{\rm int}[\sigma] & = \frac{\lambda_3}{3}\sigma^3+\frac{\lambda_4}{4}\sigma^4+\frac{\lambda_6}{6}\sigma^6\,,
\end{equation}
but kept otherwise the dimensions of the fields and coupling parameters unchanged. 

\begin{figure}[t!]
 \begin{center}
  \begin{tikzpicture}
   \node[n-style] (x1) at (128:3.18) {$x_1$};
   \node[n-style] (x2) at (90:2.5) {$x_2$};
   \draw[*-*] (x1) -- (x2);
  \end{tikzpicture}
  \hspace{4mm}
  \begin{tikzpicture}
   \node[n-style] (x) at (225:1.54) {$x$};
   \node[n-style] (x1) at (200:3.15) {$x_1$};
   \node[n-style] (x2) at (305:1.32) {$x_2$};
   \draw[*-*] (x1) -- (x);
   \draw[*-*] (x2) -- (x);
   \draw[*-*] (x) to[in=40, out=140, looseness=9] (x);
  \end{tikzpicture}
  \hspace{4mm}
  \begin{tikzpicture}
   \vspace{5mm}
   \node[n-style] (x) at (225:1.54) {$x$};
   \node[n-style] (y) at (265:1.1) {$y$};
   \node[n-style] (x1) at (230:3.15) {$x_1$};
   \node[n-style] (x2) at (290:2.6) {$x_2$};
   \draw[*-*] (x1) -- (x);
   \draw[*-*] (x2) -- (y);
   \draw[*-*] (x) to[in=90, out=100, looseness=1] (y);
   \draw[*-*] (x) to[in=-90, out=-95, looseness=1] (y);
  \end{tikzpicture}
  \end{center}
	\caption{Diagrammatic representation of the different contributions to the 2-point function Eq.~(\ref{equ:2pointFunction1D}) up to next-to-leading order in $\xi^2/A$.}
	\label{fig:2pt_1D}
\end{figure}
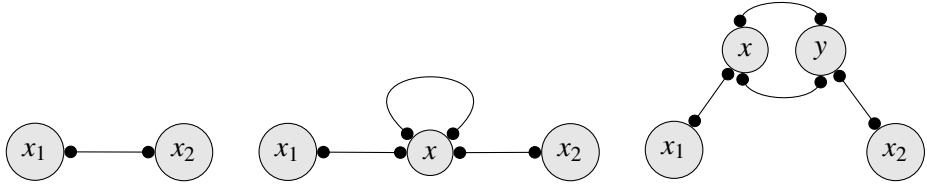

We are interested in calculating the integral of the $2$-point function over a region of size $L$
\begin{equation}
  \langle\sigma^2\rangle = \,\frac{1}{L^2}\int_L dx_1 dx_2 \,C(x_1,x_2)\, ,
 \label{equ:integralCF}
\end{equation}
which gives us the variance of the critical mode $\sigma$ for zero momentum. We expand the probability density in the $2$-point function around ${\cal P}_0$ and find 
\begin{align}
 \nonumber
  C(x_1,x_2) = & \frac{1}{{\cal Z}}\int{\cal D}\sigma \,\sigma(x_1)\sigma(x_2)\exp(-\Omega_0[\sigma]/T) \\
 \label{equ:2pointFunction1D}
  &
  \times
  \left(1-\frac{\lambda_4}{4}\frac{A}{T}\int_L{\rm d}x\,\sigma(x)^4+\frac{\lambda_3^2}{18}\frac{A^2}{T^2}\int_L{\rm d}x\int_L{\rm d}y\,\sigma(x)^3\sigma(y)^3+\dots\right) \,.
\end{align}
A diagrammatic representation of the individual contributions in Eq.~(\ref{equ:2pointFunction1D}) is shown in Fig.~\ref{fig:2pt_1D}. By evaluating Eq.~(\ref{equ:2pointFunction1D}) and integrating over the region $L$, Eq.~\eqref{equ:integralCF}, we find for the variance of the critical mode
\begin{equation}
 \langle\sigma^2\rangle = \,T\frac{\xi^2}{V}-\frac{3}{2}\left(\tilde\lambda_4-\tilde\lambda_3^2\right)T\frac{L\xi^4}{V^2}\,.
 \label{equ:Variance}
\end{equation}
In this result we have suppressed exponential factors of order $e^{-L/\xi}$. Note, that this expansion is effectively in terms of $\xi^2/A$. We find that the known, leading-order result $\propto\xi^2$ is reproduced within the approach and that the correction terms can play a potential role on the observed variance if $L\xi^2$ becomes comparable to $V$. Moreover, it can be shown that the $\lambda_6$-term in Eq.~\eqref{equ:Omega} only contributes at higher orders than considered here.

%%%%%%%%%%%%%%%%%%%%%%%%%%%%%%%%%%%%%%%%%%%%%%%%%%%%%%%%%%%%%%%%%%%%%%%%%%%
\section{Numerical results}
%%%%%%%%%%%%%%%%%%%%%%%%%%%%%%%%%%%%%%%%%%%%%%%%%%%%%%%%%%%%%%%%%%%%%%%%%%%

In order to quantify the possible impact of finite size corrections on the variance $\langle\sigma^2\rangle$ of the critical mode, we choose exemplary values for the relevant parameters. For the correlation length $\xi$ we use as input the temperature dependence of the Gaussian limit from~\cite{Nahrgang:2018afz}. This implies that the maximal equilibrium correlation length is about $3~\textrm{fm}$ near $T_c=0.15~\textrm{GeV}$ and falls off quickly to $< 1~\textrm{fm}$ below and above $T_c$. For the dimensionless, non-linear couplings we take $\tilde\lambda_3=1$ and $\tilde\lambda_4=4$. 

\begin{figure}[t]
 \centering
  \includegraphics[width=1\linewidth]{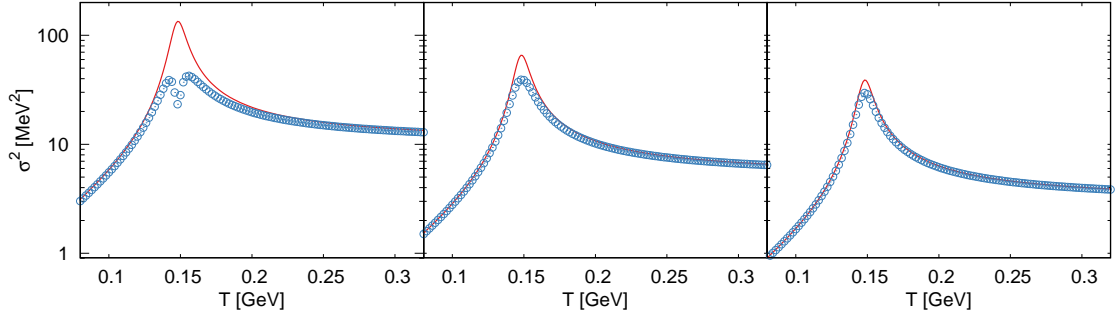}
  \caption{(Colour online) Behavior of the leading-order (red, solid lines) versus next-to-leading order results (blue circles) of the variance $\langle\sigma^2\rangle$ Eq.~(\ref{equ:Variance}) as a function of temperature $T$. The $T$-dependence of the correlation length $\xi$ is taken from~\cite{Nahrgang:2018afz}, $\tilde\lambda_3=1$, $\tilde\lambda_4=4$, $L=40~\rm{fm}$ and $V=1960~\rm{fm}^3$ (left panel). The volume increases roughly by a factor $2$ (middle panel) and $3.5$ (right panel) keeping $L$ fixed.}
 \label{fig:LOvsNLO}
\end{figure}

By varying the volume $V$ of the considered system, we may study the impact of the finite size corrections and the possible modification of the temperature dependence of the variance. This is shown in Fig.~\ref{fig:LOvsNLO}. For a fixed size $L=40~\textrm{fm}$ in one spatial direction, the leading-order result (red, solid lines) becomes smaller with increasing transverse area $A$ from $A=49~\textrm{fm}^2$ (left panel) to $A=100~\textrm{fm}^2$ (middle panel) to $A=169~\textrm{fm}^2$ (right panel), as expected. For the chosen combination of $\tilde\lambda_3$ and $\tilde\lambda_4$, the next-to-leading order corrections tend to reduce the variance $\langle\sigma^2\rangle$ even further. With increasing volume $V$, however, this effect becomes less pronounced and the leading-order result approximates Eq.~(\ref{equ:Variance}) better and better.

For a large enough $V$, the result including finite size corrections can be approximately described with the leading-order scaling behavior, but with a scaling coefficient $\langle\sigma^2\rangle\propto\xi^m$, $m<2$ (see right panel of Fig.~\ref{fig:LOvsNLO}). With decreasing volume, this scaling coefficient deviates more and more from the infinite volume expectation (see middle panel of Fig.~\ref{fig:LOvsNLO}) until the temperature dependence of the full result cannot be described anymore with a single scaling term and must be described in terms of a combination of competing scaling behaviors (see left panel of Fig.~\ref{fig:LOvsNLO}). Similar numerical observations were reported in~\cite{Nahrgang:2018afz}. We note that the particular behavior seen in Fig.~\ref{fig:LOvsNLO} is also a consequence of the fixed values chosen for the non-linear couplings $\tilde\lambda_i$. In general these may depend on the region probed in the QCD phase diagram~\cite{Bluhm:2016trm} such that a more complicated temperature dependence is to be expected. In particular an increase compared to the leading-order result is also conceivable.

%\sout{show exemplarily for given $\xi$ and parameters $\tilde\lambda_i$ a few numerical results (obtained with my Mathematica code) similar to our short net-baryon diffusion paper; show also what happens if you increase the observation volume $V_{\rm obs}$, which should go such that NLO appraoches LO result; what a $V_{\rm obs}$ must I choose to make the NLO results compatible with the LO results (with the old observation volume)}
 
%\sout{describe in detail what you observe}

%\sout{make a statement on the actual scaling behavior with $\xi$}
 
%%%%%%%%%%%%%%%%%%%%%%%%%%%%%%%%%%%%%%%%%%%%%%%%%%%%%%%%%%%%%%%%%%%%%%%%%%%
\section{Conclusion}
%%%%%%%%%%%%%%%%%%%%%%%%%%%%%%%%%%%%%%%%%%%%%%%%%%%%%%%%%%%%%%%%%%%%%%%%%%%
 
We demonstrated that for a finite size system as created in heavy-ion collisions the equilibrium expectations for the variance of critical fluctuations can substantially be modified compared to the infinite volume limit. This does not only concern the magnitude of the fluctuations but also the dependence on temperature. While in the infinite size scenario a single peak structure occurs around the critical temperature, there are values in the parameter space of the couplings for which the variance in a finite system shows a double peak structure in the vicinity of the critical point. Therefore, it can be assumed that already the equilibrium shape (i.e. without even taking dynamical effects into account) of higher-order cumulants, which will be treated explicitly elsewhere, can be altered significantly in a realistic situation. A qualitatively similar behavior was observed in the recent dynamical approach \cite{Nahrgang:2018afz}.

We note here, that the pursued method of integrating the $2$-point correlation function results in a somewhat unexpected expansion parameter, which is $\xi^2/A$ instead of the so far reported $\xi^3/V$, cf.~\cite{Stephanov:2008qz}. It would be interesting if experimentally the $n$-point correlation functions could be analyzed in addition to the integrated cumulants.

%\sout{main observation, main points}
 
%\sout{only variance studied here, higher-order cumulants reported in detail in a forthcoming work} 
 
%\sout{qualitatively similar behavior (also with $V_{\rm obs}$) seen in a dynamical approach (our short paper on stochastic baryon diffusion)}
 
%\sout{be careful of what you show and how deep you go into the subject}

%%%%%%%%%%%%%%%%%%%%%%%%%%%%%%%%%%%%%%%%%%%%%%%%%%%%%%%%%%%%%%%%%%%%%%%%%%%
\section*{Acknowledgments}
%%%%%%%%%%%%%%%%%%%%%%%%%%%%%%%%%%%%%%%%%%%%%%%%%%%%%%%%%%%%%%%%%%%%%%%%%%%

The authors acknowledge the support by the program ``Etoiles montantes en Pays de la Loire 2017''. M.~Bluhm also acknowledges the partial support by the European Union's Horizon 2020 research and innovation program under the Marie Sk\l{}odowska Curie grant agreement No 665778 via the National Science Center, Poland, under grant Polonez UMO-2016/21/P/ST2/04035. The authors thank T.~Sch\"afer for stimulating discussions.

%%%%%%%%%%%%%%%%%%%%%%%%%%%%%%%%%%%%%%%%%%%%%%%%%%%%%%%%%%%%%%%%%%%%%%%%%%%

\end{document}